\documentstyle[12pt]{article}
\begin{document}
\title{Gravitational radiation, vorticity and super--energy: A conspicuous threesome}
\author{L. Herrera$^1$\thanks{e-mail: lherrera@usal.es}\\
{$^1$Instituto Universitario de F\'isica
Fundamental y Matem\'aticas},\\ {Universidad de Salamanca, Salamanca 37007, Spain}}
\maketitle

\abstract{We elaborate on  the  link relating gravitational radiation, vorticity and a flux of super--energy on the plane orthogonal to the vorticity vector. We  examine the vorticity appearing in the congruence of  observers at the outside of the source, as well as the vorticity of the fluid distribution, the source of the gravitational radiation is made of. The information provided by the study of the physical aspects of the source  poses new questions  which  could, in principle, be solved by the observational evidence.  Besides the study of the  theoretical issues associated to such relationship, we also stress  the new observational possibilities to detect gravitational radiation, appearing  as consequence of  the above mentioned link.   The high degree of development achieved in  the gyroscope   technology, as well as recent proposals   to detect rotations by means of ring lasers,  atom interferometers, atom lasers  and  anomalous spin--precession experiments, lead us to believe that an alternative to the laser interferometers used so far to detect gravitational waves, may be implemented based on the detection of the vorticity associated with gravitational radiation. Additionally, this kind of detectors  might be able to elucidate the open question about the physical properties of the tail of the waves appearing as the consequence of the violation of the  Huygens's principle in general relativity.} 

\section{Introduction}

The detection of gravitational radiation by means of  laser  interferometers represents a major breakthrough in the development of the theory of gravitation \cite{1}. However, the  apparatus involved in such experiments is extremely expensive and  requires the participation of  large number of researchers,  as a consequence of which this observational approach is accessible only to huge collaboration groups endowed with very large budgets. Therefore it is clearly alluring the idea to  detect gravitational radiation by means of experiments  based on effects, different from those involved in laser interferometers,  that could be carried out by means of a cheaper technology and  thereby affordable to a larger number of groups of research all around the world.

Our  purpose in this paper is twofold. On the one hand we shall  present a comprehensive discussion about  the link between gravitational radiation, vorticity and  a flux of super--energy  on the plane orthogonal to the vorticity. On the other hand we would like to call the attention to the  potential observational  application  of this effect on the detection of gravitational radiation.

The relationship between gravitational radiation and vorticity  was put forward for the first time in Reference \cite{2}, using the Bondi approach \cite{3, 4}. It was subsequently discussed  by many researchers (see for example References \cite{5, 6, 7, 8, 9, 10, 11} and references therein). Short afterwards, it was established  that the link between gravitational radiation and vorticity becomes intelligible if we introduce another element into the discussion, namely a flux of super--energy on the plane orthogonal to the vorticity vector \cite{8, 10}. 

The idea to relate the vorticity to a  flow of super--energy, put forward in Reference \cite{8}, stems~from an intriguing fact related to  the space--time generated by a  magnetic dipole plus a central charge. The~point is that such  space--time  appears to be  stationary (non--static), that is, there is a vorticity  in the congruence of  observers at rest with respect to the source. In order to explain this strange situation, Bonnor \cite{8N} invokes a curious result  of classical electrodynamics concerning the Poynting vector of a charged magnetic dipole (see Reference \cite{N9}). Indeed, it appears that for a charged magnetic dipole there is a  non--vanishing flow of electromagnetic energy (as described by the Poynting vector), in spite of the fact that the system is time--independent. Thus, Bonnor explains  the appearance of vorticity in such  a system, as produced by the circular flow of electromagnetic energy (later on it was established that this is the explanation for the appearance of vorticity in any stationary electrovacuum solution to the Einstein equations \cite{N10}). In Reference  \cite{8} we have just extended the Bonnor's idea by replacing the electromagnetic energy by super--energy, in order to explain the appearance of vorticity associated to the emission of gravitational radiation.

All the discussion  above concerns the vorticity of the congruence of observers in the vacuum (outside the source). However a similar link exists between gravitational radiation and the vorticity of the congruence of the fluid elements forming the source.  The combined discussion  from the point of view of  observers, both,  outside and inside the source, provides a deeper insight  into the problem and poses new questions which eventually could be answered by experimental observation.

Before entering into the core of the discussion it would be convenient  to specify in some detail the three  essential ingredients  of our proposal.

%%%%%%%%%%%%%%%%%%%%%%%%%%%%%%%%%%%%%%%%%%
\section{Vorticity, Gravitational Radiation and Super--Poynting Vector}

Let us start with the   concept of  vorticity of a congruence.  As is well known  the vorticity vector {${\bf \Omega}$}  defining   the vorticity  of a congruence, 
describes the rate of rotation (the proper angular velocity) with respect to  proper time at any reference particle at rest in the rotating frame, relative to the local compass of inertia. This compass of inertia is physically realized by three gyroscopes spinning about three orthogonal axes. It is then obvious that  {$-{\bf \Omega}$} describes the rotation of the compass of inertia (gyroscope) with respect to reference particles.  From these comments the  suitability  of gyroscopes to detect rotations becomes evident.  %bold is necessary? Please uniform its expressions in the main text. such as bold or normal.

Two very important relativistic effects can be identified and measured  by  means of gyroscopes. One is the well known Fokker--de Sitter effect,  which refers to the precession of a gyroscope following a closed orbit around a spherically symmetric mass distribution (a version  of Thomas precession in the Schwarzschild spacetime).  This effect can be verified with a great degree of accuracy by observing the rotation of the earth--moon system  around the sun \cite{12}. The other effect is the Lense--Thirring--Shiff precession, which is  much more difficult to observe and refers to the generation of vorticity in the congruence of observers due to the rotation of the source (the so called  frame dragging effect,  see~ Reference \cite{13} for a detailed discussion on this issue). The  Gravity Probe B experiment not only confirmed the reality of the  Lense--Thirring--Shiff   precession \cite{14, 15} but also put in evidence the high degree of development achieved so far in the gyroscope technology.

Next, let us say few words about super--energy and super--Poynting vector.

In any classical field theory, energy is a quantity
defined in terms of potentials and their first derivatives. Accordingly, in
general relativity a definition of energy through a tensor variable is impossible since, as is well known,  we cannot form a tensor with the metric and its first derivatives alone. In other words, a local description of
gravitational energy in terms of true invariants (tensors of any
rank) is  not possible within the context of the theory. This situation has lead researchers to look for non--local definitions of energy or to describe it by means of  pseudo--tensors or to resort 
to  a succedaneous definition of energy.

Super--energy is an example of this last alternative.  It may be defined
from the Bel  or the  Bel--Robinson tensor  \cite{16, 17, 18}
(they both coincide in vacuum) and has been shown to be very useful
when it comes to explaining  a number of  phenomena in the context of
general relativity.

The Bel and the Bel--Robinson tensors are obtained from the Riemann and the Weyl tensor (as well as their dual) respectively,  by analogy  with  the form in which the energy--momentum tensor of the electromagnetic field depends on the Maxwell tensor (and its dual). Obviously they coincide in vacuum but differ  within the source. Also, by analogy with electromagnetism, a super--Poynting vector describing the flux of super--energy can be defined. This quantity is particularly relevant for our discussion since in the theory of  the super--Poynting vector, a state of gravitational radiation is associated to a  non--vanishing component of the latter (see References \cite{16, 17, 18}).  This is in agreement with the established link between the super--Poynting vector and the news functions \cite{8}, in the context of the Bondi--Sachs approach \cite{3, 4} (see below). 

Finally, we shall provide a summarized description of gravitational radiation. This will be done for, both, the space--time outside the source, which we consider bounded in order for the spacetime to be asymptotically flat,  as well as for the space--time  within the source.The  following two subsections are devoted to these issues.

\subsection{Gravitational Radiation in An Asymptotically Flat Vacuum Space--Time: The Bondi Approach.}
\unskip

The Bondi formalism \cite{3}  is centered around  the general form of an axially and reflection symmetric asymptotically flat
metric given by  (for the general case see Reference \cite{4})
\begin{equation}
\begin{array}{cll}
ds^2 & = & \left(\frac{V}{r} e^{2\beta} - U^2 r^2 e^{2\gamma}\right) du^2
+ 2 e^{2\beta} du dr \\
& + & 2 U r^2 e^{2\gamma} du d\theta
- r^2 \left(e^{2 \gamma} d\theta^2 + e^{-2\gamma} \sin^2{\theta} 
d\phi^2\right),
\label{Bm}
\end{array}
\end{equation}
where $V, \beta, U$ and $\gamma$ are functions of
$u, r$ and $\theta$.

The coordinates are numbered $x^{0,1,2,3} = u, r, \theta, \phi$, respectively.
$u$  is a timelike coordinate ($g_{uu}>0$~) and the hypersurfaces  {$u=constant$} define  null surfaces (their normal vectors are  null vectors), which~in the null infinity ($r\rightarrow \infty$) coincide with the Minkowski null light cone
open to the future. $r$ is a null coordinate ($g_{rr}=0$) and $\theta$ and
$\phi$ are two angle coordinates (see Reference \cite{3} for details).%should constant be normal? 

Regularity conditions in the neighborhood of the polar axis
($\sin{\theta}=0$), imply that
as $\sin{\theta} \rightarrow 0$
\begin{equation}
V, \beta, U/\sin{\theta}, \gamma/\sin^2{\theta},
\label{regularity}
\end{equation}
each equals a function of $\cos{\theta}$ regular on the polar axis.

Expanding the metric functions in series of $1/r$,
and using the field equations one  gets

\begin{equation}
\gamma = c r^{-1} + \left(C - \frac{1}{6} c^3\right) r^{-3}
+ {\mathcal{O}}(r^{-n})\quad  n\geq 4,\nonumber
\label{ga}
\end{equation}

\begin{equation}
U = - \left(c_\theta + 2 c \cot{\theta}\right) r^{-2} + \left[2
N+3cc_{\theta}+4c^2 \cot{\theta}\right]r^{-3}+ {\mathcal{O}}(r^{-n}) \quad  n\geq 4,
\label{U}
\end{equation}

\begin{eqnarray}
V  = r - 2 M
 -  \left( N_\theta + N \cot{\theta} -
c_{\theta}^{2} - 4 c c_{\theta} \cot{\theta} -
\frac{1}{2} c^2 (1 + 8 \cot^2{\theta})\right) r^{-1} + {\mathcal{O}}(r^{-n})\quad  n\geq 2,\nonumber
\label{V}
\end{eqnarray}

\begin{equation}
\beta = - \frac{1}{4} c^2 r^{-2} + {\mathcal{O}}(r^{-n})\quad  n\geq 3,\nonumber
\label{be}
\end{equation}
 where letters as subscripts denote derivatives and $c$, $C$, $N$ and $M$ are functions of $u$ and $\theta$ satisfying the~constraint
\begin{equation}
4C_u = 2 c^2 c_u + 2 c M + N \cot{\theta} - N_\theta.
\label{C}
\end{equation}

The three functions $c, M$ and $N$ are further
constrained  by the supplementary conditions
\begin{equation}
M_u = - c_u^2 + \frac{1}{2}
\left(c_{\theta\theta} + 3 c_{\theta} \cot{\theta} - 2 c\right)_u,
\label{Mass}
\end{equation}
\begin{equation}
- 3 N_u = M_\theta + 3 c c_{u\theta} + 4 c c_u \cot{\theta} + c_u c_\theta.
\label{N}
\end{equation}

In the static case $M$ is related to the mass of the system and was called by Bondi the ``mass aspect'', whereas $N$ and $C$
are closely related to the dipole and quadrupole moment respectively.

Using the mass aspect,  Bondi defines the mass $m(u)$ of the system as
\begin{equation}
m(u) = \frac{1}{2} \int_0^\pi{M \sin{\theta} d\theta},
\label{m}
\end{equation}
which, using  (\ref{Mass}) and (\ref{regularity}) produces
\begin{equation}
m_u = - \frac{1}{2} \int_0^\pi{c_u^2 \sin{\theta} d\theta}.
\label{muI}
\end{equation}

We may now summarize the more relevant results emerging  from the Bondi approach:

\begin{itemize}
\item If $\gamma, M$ and $N$ are known  on some null hypersurface $u=a $(constant) and
$c_u$ (the news function) is known for all $u$ in the interval
$a \leq u \leq b$,
then the system is fully determined in that interval. This implies that
whatever happens at the source, leading to changes in the field,
it can only do so by affecting $c_u$ and viceversa. This result  establishes in an unmistakable way  the relationship between news function
and the occurrence of radiation.
\item As it follows from (\ref{muI}), the mass of a system is constant
if and only if there are no news.
\end{itemize}

Now, for an observer at rest in the frame of (\ref{Bm}), the four-velocity
vector has components
\begin{equation}
V^{\alpha} = \left(\frac{1}{A}, 0, 0, 0\right),
\label{fvct}
\end{equation}
with
\begin{equation}
A \equiv \left(\frac{V}{r} e^{2\beta} - U^2 r^2 e^{2\gamma}\right)^{1/2}.
\label{A}
\end{equation}

For the observer defined by  (\ref{fvct}) the vorticity vector may be
written as (see Reference \cite{2} for details)
\begin{equation}
\omega^\alpha = \left(0, 0, 0, \omega^{\phi}\right),
\label{oma}
\end{equation}
while for its absolute value we obtain
\begin{equation}
\begin{array}{cll}
\Omega  &\equiv & \left(- \omega_\alpha \omega^\alpha\right)^{1/2} = -\frac{1}{2r} ( c_{u \theta}+2 c_u \cot \theta)  \\ &+&\frac 1{r^2} \left[ M_{\theta}-M (c_{u \theta}+2 c_u \cot 
\theta)-c c_{u
\theta}+6 c c_u \cot \theta+2 c_u c_{\theta} \right] +{\mathcal{O}}(r^{-n})\quad  n\geq 3.
\label{Om2}
\end{array}
\end{equation}

The first term on the right hand side of  (\ref{Om2}) describes the contribution of gravitational radiation to vorticity, it vanishes if and only if there are no news, that is, if there is no gravitational radiation. Instead, the second term  survives even if $c_u=0$. In this latter case the vorticity has to be related  to the tail of the wave associated to the $M_\theta$ term. We shall discuss further on this issue in the last section.

Finally, we shall need an expression for the super--Poynting vector.
The  super--Poynting vector  based on the Bel--Robinson tensor is
defined
as (see Reference \cite{8} for details)
\begin{equation}
P_{\alpha}=\eta_{\alpha \beta \gamma
\delta}E^{\beta}_{\rho}H^{\gamma
\rho}u^{\delta},
\label{p1}
\end{equation}
where $\eta_{\alpha \beta \gamma
\delta}$ is the permutation symbol and 
$E_{\mu\nu}$ and $H_{\mu\nu}$, are the electric and magnetic parts of
Weyl tensor, respectively. They are defined from the Weyl tensor  $C_{\alpha
\beta \gamma
\delta}$ and its dual
$\tilde C_{\alpha \beta \gamma \delta}$ by
contraction with the four
velocity vector, as
\begin{equation}
E_{\alpha \beta}=C_{\alpha \gamma \beta
\delta}V^{\gamma}V^{\delta},
\label{electric}
\end{equation}
\begin{equation}
H_{\alpha \beta}=\tilde C_{\alpha \gamma \beta
\delta}u^{\gamma}u^{\delta}=
\frac{1}{2}\eta_{\alpha \gamma \epsilon
\delta} C^{\epsilon
\delta}_{\quad \beta \rho}
V^{\gamma}
V^{\rho}.
\label{magnetic}
\end{equation}

For the Bondi metric we obtain 
\begin{equation}
P_\mu=(0,P_r,P_\theta, 0),
\end{equation}
where the leading terms of each  component are
\begin{equation}
P_r=-\frac{2c_{uu}^2}{r^2} + \mathcal{O}(r^{-n})\quad  n\geq 3,
\label{point1}
\end{equation} 
\begin{eqnarray}
P_\theta =-\frac{2}{r^2 \sin\theta}\left[
(2c_{uu}^2c
+c_{uu}c_{u})\cos\theta +\left(c_{uu}c_{\theta
u}+c_{uu}^2c_{\theta}\right)\sin\theta \right]+ {\mathcal{O}}(r^{-n})\quad  n\geq 3
.\label{p2}
\end{eqnarray}

Thus, we have a radial component describing the propagation of super--energy along  the generators of the null hypersurface $u=constant$ and a ``meridional'' component which is the one related to the vorticity (see References \cite{8, 10} for details).  A striking confirmation of the link between vorticity and and a flow of super--energy on the the plane orthogonal to the vorticity vector  is provided by analyzing the Einstein--Rosen  metric. In this case there is a radial component of the Super--Poynting vector related to the propagation of the gravitational wave, whereas the absence of vorticity is explained by the fact that  there is no  flow of super--energy in any plane of the 3-space (see References \cite{10} for details).The vanishing of the $\phi$-component is due to the reflection symmetry of the Bondi metric.

We shall next provide a brief description of the the source of the gravitational radiation.
\subsection{Gravitational Radiation within the Source}
The Bondi approach sketched in the previous subsection is very powerful to analyze the radiative space--time outside the source but fails as we approach the source, due to the appearance of caustics,  and of course is not suitable for describing the situation within the source. Accordingly in order to describe the situation within the fluid distribution we shall resort to a different framework, developed in Reference \cite{19} and based on the $1+3$  approach \cite{20, 21, 22, 23}, whose main characteristics are described below (for details see Reference \cite{19}).

Thus, let us  consider,  axially (and reflection) symmetric sources,  for which  the most general line element may be written  as:
\begin{equation}
ds^2=-A^2 dt^2 + B^2 \left(dr^2
+r^2d\theta^2\right)+C^2d\phi^2+2Gd\theta dt, \label{1b}
\end{equation}
where $A, B, C, G$ are positive functions of $t$, $r$ and $\theta$. We number the coordinates $x^0=t, x^1=r, x^2= \theta, x^3=\phi$.

The source is filled with an anisotropic and dissipative fluid, therefore the energy momentum tensor may be written in the ``canonical'' form, as 
\begin{equation}
{T}_{\alpha\beta}= (\mu+P) V_\alpha V_\beta+P g _{\alpha \beta} +\Pi_{\alpha \beta}+q_\alpha V_\beta+q_\beta V_\alpha,
\label{6bis}
\end{equation}
where  $\mu$ is the energy
density, $q_\alpha$ is the  heat flux, whereas  $P$ is the isotropic pressure and $\Pi_{\alpha \beta}$ is the anisotropic tensor. We are considering an Eckart frame  where fluid elements are at rest.

For the line element (\ref{1b}), it can be shown that the heat flux vector is determined by two scalar functions, whereas  three scalar functions describe the anisotropic tensor.  Also, there is a vorticity vector, with a single component along the $\phi$ direction, defined in terms of a single scalar function $\Omega$ given by
\begin{equation}
\Omega =\frac{G(\frac{G_{,r}}{G}-\frac{2A_{,r}}{A})}{2B\sqrt{A^2B^2r^2+G^2}}.
\label{no}
\end{equation}

Regularity conditions at the centre imply that: $G=0\Leftrightarrow \Omega=0$.

Finally we shall need to consider the super--Poynting vector $P^\mu$ within the source. As we know from Reference \cite{8}, there is always a non-vanishing component of $P^\mu$, on the
plane orthogonal to a unit vector along which there is a non-vanishing component of vorticity (the $\theta-r$ plane).
Instead, $P^\mu$~vanishes along the $\phi$-direction since there are no motions along this latter direction, because of the reflection symmetry. 

Explicit expressions  for the components of  the super--Poynting vector  may be found in Reference~\cite{19}. Suffice is to say at this point that we can identify three different contributions in the super--Poynting vector. On the one hand we have contributions from the  heat transport process. These are in principle independent of the magnetic part of the Weyl tensor, which explains why (some of them ) remain in the spherically symmetric limit.  On the other hand  we have contributions from the magnetic part of the Weyl tensor. These are of two kinds. On the one hand, contributions associated with the propagation of gravitational radiation within the fluid and on the other hand, contributions of the flow of super--energy associated with the vorticity on the plane orthogonal to the direction of propagation of the radiation, this is the effect relevant for our discussion here. Both contributions are intertwined and it appears  to be impossible to disentangle them  through two independent scalars.

Before concluding this section, the following remarks are in order:
\begin{itemize}
\item The gravitational radiation being a dissipative process,  we should expect that an entropy generator factor be present in the source of radiation. This has been discussed with some detail in Reference~\cite{24}.
\item As a consequence of the previous point, the exterior of a gravitationally radiating source is necessarily filled with a null fluid, produced by the dissipative processes inherent to the emission of gravitational radiation. In other words the assumption of vacuum in the Bondi approach has to be regarded as an approximation, see Reference \cite{25} for a discussion on this issue.
\item It is important to keep in mind the difference between the steady vorticity of the stationary case (e.g., the one of the Kerr metric) and the vorticity considered here. In the former case there is no gravitational radiation, although the vorticity is also related to a flux of super--energy on the plane orthogonal to the vorticity vector (see Reference \cite{27} for details).
\item It is important to stress that we are dealing here, exclusively,  with sources of gravitational radiation  represented by a fluid distribution. In other words the emission of gravitational radiation is entirely  due to changes in their relativistic multipole  moments. Accordingly,  we are excluding  gravitational radiation of the ``synchrotron'' type produced by accelerated massive particles or the two body problem.

\end{itemize}

We have now available all elements we need for our discussion.

\section{Discussion}
The main point to retain from the preceding sections is that whenever gravitational radiation is emitted we should expect the appearance of vorticity in the congruence of the world lines of observers. Accordingly, any experimental device intended to measure  rotations could be  a potential detector of gravitational radiation as well. Of course, extremely high sensitivities have to be reached, for these detectors to be operational. However, even if the present technology might not be up to the required sensitivities,  the  intense activity deployed in recent years in this field,  invoking  ring lasers,  atom interferometers, atom lasers,  anomalous spin--precession, trapped atoms  and  quantum interference (see References \cite{28, 29, 30, 31, 32, 34, 35, 36, 37, 38, 39, 40} and references therein), besides the incredible sensitivities obtained so far  in  gyroscope technology and exhibited in the Gravity Probe B experiment \cite{14, 15}, make us confident in that this kind of detectors may be operating in the foreseeable future.

As already mentioned in the Introduction, the association of gravitational radiation and vorticity was first put in evidence from the study of the space--time outside the source. However, and this is perhaps one of our main points in this work, the information provided by the inclusion of the physical properties of the source into the whole picture, leads to  new fundamental questions which could be answered by the observational evidence. 

Indeed, in Reference \cite{26} an exhaustive analysis of axially symmetric fluid distributions  just after its departure from equilibrium, has been carried out  at the smallest time scale at which we can detect signs of dynamical evolution. It was then obtained that  the departure from equilibrium and the ensuing  evolution of all variables, is controlled by a single function—called the fluid news function—in analogy with the Bondi's news function. Such a function is  related  to  the time derivative of the vorticity vector, putting in evidence  the link between vorticity and radiation within the source.

However, the most  relevant result obtained from this study, concerning  the present discussion,  is  the fact that  both the magnetic part of the Weyl tensor and $\Omega$ vanish at the time scale under consideration, whereas their first time derivatives are non--vanishing   at that same time scale, thereby suggesting that both phenomena (radiation and vorticity), as well as the non--vanishing component of the super--Poynting vector on the plane orthogonal to the vorticity, occur essentially simultaneously.  This is at variance with the point of view assumed  in Reference \cite{8} where   it was  assumed that radiation precedes the appearance of vorticity.  Although this point deserves further discussion from the theoretical point of view,  we hope that it could eventually be elucidated by experimental observation.

Another important issue appearing from the study of the physical  properties of the source, concerns the tail of the gravitational wave. Indeed, as  it is apparent  from (\ref{Om2}), the vorticity related to gravitational radiation combines two different type of contributions. On the one hand we have the term of order ${\mathcal{O}}(r^{-1})$ which is directly related to the non--vanishing of the news function, that is, it~is the vorticity associated to the emission of gravitational radiation. On the other hand, the term of  ${\mathcal{O}}(r^{-2})$ does not vanish after the emission of gravitational radiation stops ($c_u=0$) and therefore has to be related to the tail of the wave, appearing as consequence of the violation of the Huygens's principle in curved space--times.  Therefore the transition from a radiating regime to the static one seems to be  forbidden and may happen only asymptotically in time.  However, when we take into account the source of the gravitational radiation, it appears that the above conclusion is not evident. Let us analyze this  issue  more closely.

The appearance of tails after the emission of gravitational radiation has been established in studies considering exclusively the space--time outside the source and as a matter of fact far from the source (see References \cite{41, 42, 43, 44, 45, 46, 47} and references therein). However, a recent study on the transition  of a gravitationally radiating fluid source to equilibrium \cite{48}, shows that such a transition may take place at time scale of the order of thermal relaxation time, thermal adjustment time or hydrostatic time (whichever is larger). The explanation for such disagreement  might be found in the fact that in the studies carried on in the vacuum, some physical phenomena describing the interaction of the field with the source might have been overlooked. This result strengthens further the importance to analyze the problem, from both the outside and the inside of the source. At any rate it is clear that observational detection of the ${\mathcal{O}}(r^{-2}) $ term in the vorticity, would help to clarify this point.

Finally, the following remark is in order: The presented discussion  was carried out in the context of general relativity, however, as is known,   due to several astronomical observations at different scales, which pose some problems of interpretation within the context of the ``classical'' Einstein theory, some researchers have found it useful to modified the general relativity in order to accommodate the above mentioned observational data. The interest on such alternative theories is therefore fully justified and the obvious question arises: how would the presented results change if, instead of using GR, we use any of the alternative theories? Although a thorough answer to the above question is out of the scope of this work, it seems very likely that the additional terms corresponding to these theories should appear in the expression for the vorticity and therefore the observational evidence could help to  discriminate between different alternatives. 

\vspace{12pt} 

%=====================================

% The following MDPI journals use author-date citation: Arts, Econometrics, Economies, Genealogy, Humanities, IJFS, JRFM, Laws, Religions, Risks, Social Sciences. For those journals, please follow the formatting guidelines on http://www.mdpi.com/authors/references
% To cite two works by the same author: \citeauthor{ref-journal-1a} (\citeyear{ref-journal-1a}, \citeyear{ref-journal-1b}). This produces: Whittaker (1967, 1975)
% To cite two works by the same author with specific pages: \citeauthor{ref-journal-3a} (\citeyear{ref-journal-3a}, p. 328; \citeyear{ref-journal-3b}, p.475). This produces: Wong (1999, p. 328; 2000, p. 475)

\end{document}